\documentstyle[epsf]{article}

\begin{document}

\title{\null\vspace*{-2truecm}\hfill\mbox{\small IISc-CTS-6/00}\\
            \vspace*{-0.5truecm}\hfill\mbox{\tt\small hep-ph/0003192}\\
            Constraints on $m_s$ and $\epsilon'/\epsilon$ from Lattice QCD}
\author{Apoorva Patel\\
       CTS and SERC, Indian Institute of Science, Bangalore-560012, India\\
       E-mail: adpatel@cts.iisc.ernet.in}
\date{\small PACS Nos.: 12.38.Gc, 12.15.Ff, 13.20.Eb.}
\maketitle

\vspace{-0.5truecm}
\noindent{Keywords: Lattice QCD, Chiral Perturbation Theory, QCD Sum Rules,
          Strange Quark Mass, CP Violation, Hadronic Matrix Elements,
          Operator Product Expansion, Final State Interactions.}

\medskip
\centerline{\small\sl Invited talk presented at the Workshop on High
                      Energy Physics Phenomenology ``WHEPP-6'',}
\centerline{\small\sl 3-15 January 2000, Institute of Mathematical
                      Sciences, Chennai. To appear in the proceedings.}

\begin{abstract}
Results for light quark masses obtained from lattice QCD simulations are
compared and contrasted with other determinations. Relevance of these
results to estimates of $\epsilon'/\epsilon$ is discussed.
\end{abstract}

\vspace*{-1mm}
\section{Determination of Light Quark Masses}
Quark masses are not physical observables in QCD, rather they enter the
theory as parameters in the Lagrangian. Their values depend on the QCD
renormalisation scale, and three quantitative approaches have been used
to determine them.

\vspace*{-1mm}
\subsection{Chiral perturbation theory}
This is a low energy ($E\leq1~{\rm GeV}$) effective field theory of QCD in
presence of spontaneous chiral symmetry breaking. With $m_q \ll \Lambda_{QCD}$,
the pseudo-Nambu-Goldstone bosons are the dominant fields at low energy.
$O(p^2,m_q)$ terms in the effective Lagrangian are fixed by the spectrum.
$O(p^4,p^2m_q,m_q^2)$ terms are estimated using resonance saturation and
large $N_c$ power counting, as well as from phenomenological fits to various
form factors. Small electromagnetic and isospin breaking effects are
systematically included, and bounds are obtained on quark mass ratios.

The range of validity of this effective field theory expansion cannot be
convincingly established. Also the absolute mass scale has to be fixed
from experimental data. Yet renormalisation group invariant dimensionless
mass ratios can be tightly constrained \cite{leutwyler}.
\begin{equation}
{m_u \over m_d} ~=~ 0.553 \pm 0.043 ~,~
{m_s \over m_d} ~=~ 18.9 \pm 0.8 ~,~
{m_s \over m_u} ~=~ 34.4 \pm 3.7 ~,~
\end{equation}
\begin{equation}
{2m_s \over m_u+m_d} ~\equiv~ {m_s \over m_{ud}} ~=~ 24.4 \pm 1.5 ~,~
{m_s - m_{ud} \over m_d-m_u} ~=~ 40.8 \pm 3.2 ~.
\end{equation}
These mass ratios are typically combined with $m_s(1~{\rm GeV})$ extracted
from QCD sum rules to give individual quark masses.

\vspace*{-1mm}
\subsection{QCD sum rules}

In this approach, two-point hadronic current correlators are evaluated
using the Operator Product Expansion (OPE) and perturbative QCD in the
Euclidean region $q^2<0$. The expansion in inverse powers of $-q^2$ is
then analytically continued to the physical region $q^2>0$, and matched
to the experimental data using dispersion relations. The experimental data
is organised in terms of contributions from well-known leading poles and
subleading branch-cuts from the QCD continuum. Subtractions are used in
the dispersion relations to suppress large continuum contributions, while
threshold factors and moments help in suppressing contribution of regions
near the poles where perturbative QCD is inapplicable. The results
critically depend on proper choice of the boundary conditions.

Pseudoscalar, scalar, $e^+e^--$annihilation and $\tau-$decay sum rules have
been used by various groups to extract $m_s$. In the $\overline{MS}$ scheme,
the required $\beta$ and $\gamma$ functions are known to $4-$loop precision,
giving the most accurate results \cite{narison}:
\begin{equation}
\overline{m}_s (1~{\rm GeV}) = 162.5 \pm 15.5~{\rm MeV} ~,~
\overline{m}_s (2~{\rm GeV}) = 117.8 \pm 12.3~{\rm MeV} ~.
\end{equation}
Analysis of Cabbibo-suppressed $\tau-$decay gives
$\overline{m}_s (M_\tau) = 119 \pm 24$ MeV \cite{pichprades}.

The major uncertainty in the sum rule approach is the asymptotic nature of
the perturbative QCD expansion---in the longitudinal channel it cannot be
pushed beyond $O(\alpha_s^3)$.

\vspace*{-1mm}
\subsection{Lattice QCD simulations}

These provide a first principle non-perturbative determination of quark
masses without additional assumptions. It is convenient to convert the
running quark masses to renormalisation group invariant quark masses,
completely within the lattice regularisation framework.
\begin{equation}
m_q^{RGI} ~=~ \lim_{\mu\rightarrow\infty}
	      \{ (2\beta_0g^2(\mu))^{-\gamma_0/2\beta_0} m_q(\mu)\} ~.
\end{equation}
With improvements in simulation algorithms and computers, the systematic
errors from finite lattice spacing, finite lattice size and the quenched
approximation are gradually coming under control. Most simulations are
in the range $m_s/3 \leq m_q \leq 3 m_s$, and $m_u = m_d$ is assumed.
Results from several simulations are combined to extrapolate to
$\mu\rightarrow\infty$.

Perturbative relations connecting lattice and $\overline{MS}$ values show
that including higher order corrections decreases $\overline{m}_q$. Moreover,
at $\mu=2~{\rm GeV}$, the $N^3LO$ correction is comparable to the $N^2LO$
one, and decreases $\overline{m}_q$ by about $10\%$ \cite{chetyrkin}.
A rigorous comparison of lattice and $\overline{MS}$ values should therefore
be made at $\mu > 2~{\rm GeV}$, contrary to common practice.

\begin{figure}[t]
\vspace{-9mm}
\begin{center}
\leavevmode
\epsfxsize=7.5cm
\epsfbox{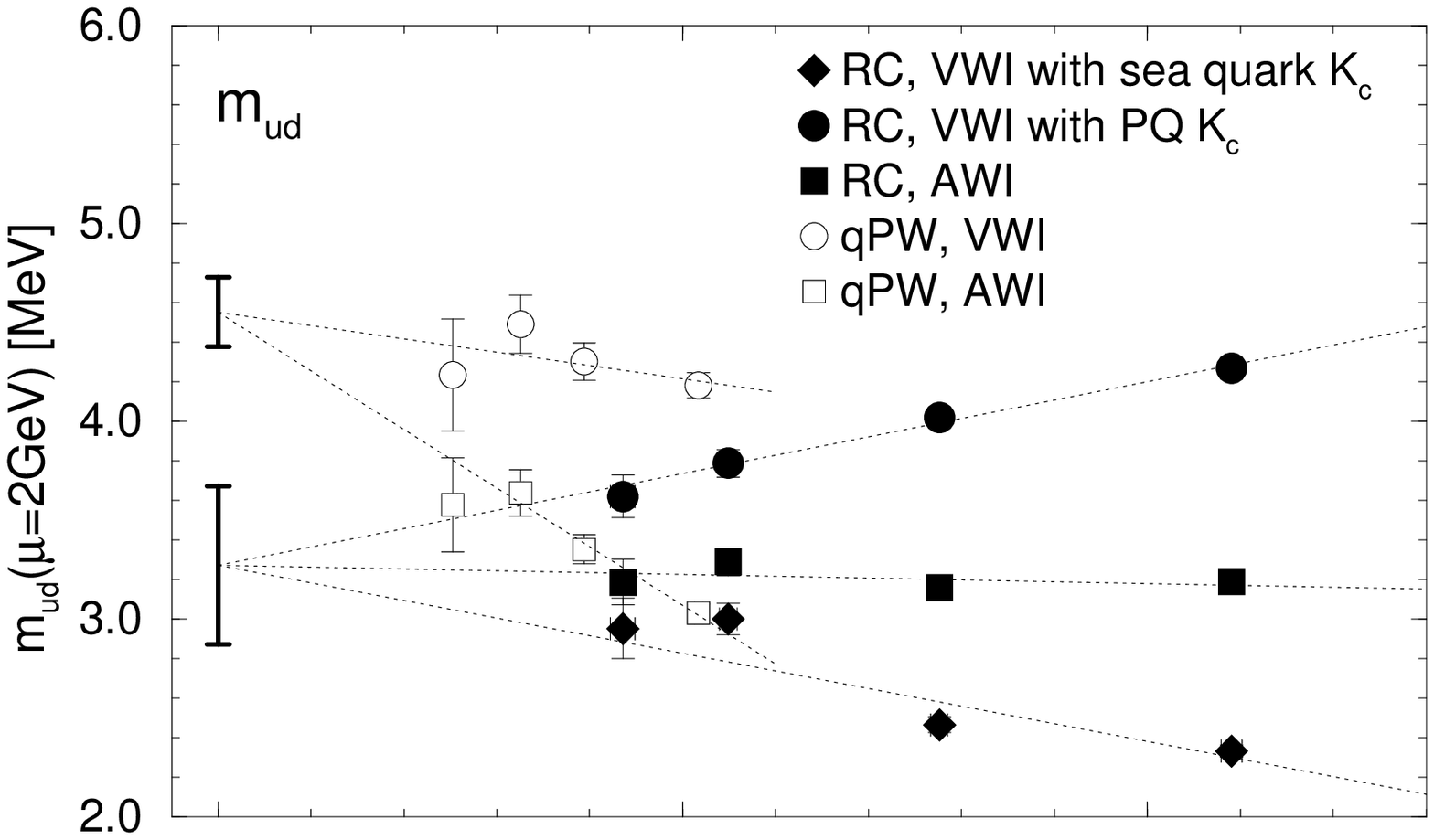}
\vspace{-26mm}
\leavevmode
\epsfxsize=7.5cm
\epsfbox{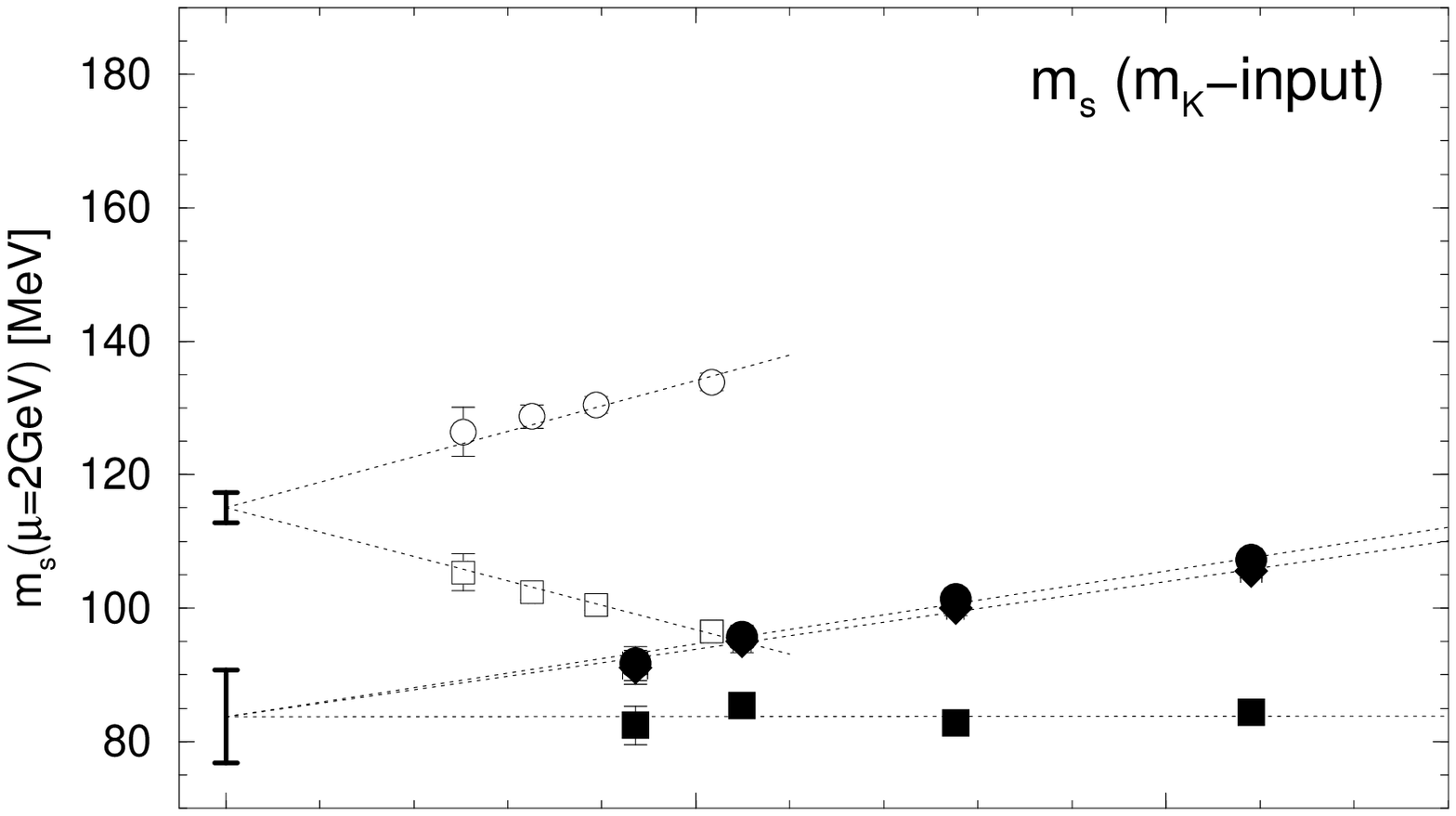}
\vspace{-26mm}
\leavevmode
\epsfxsize=7.5cm
\epsfbox{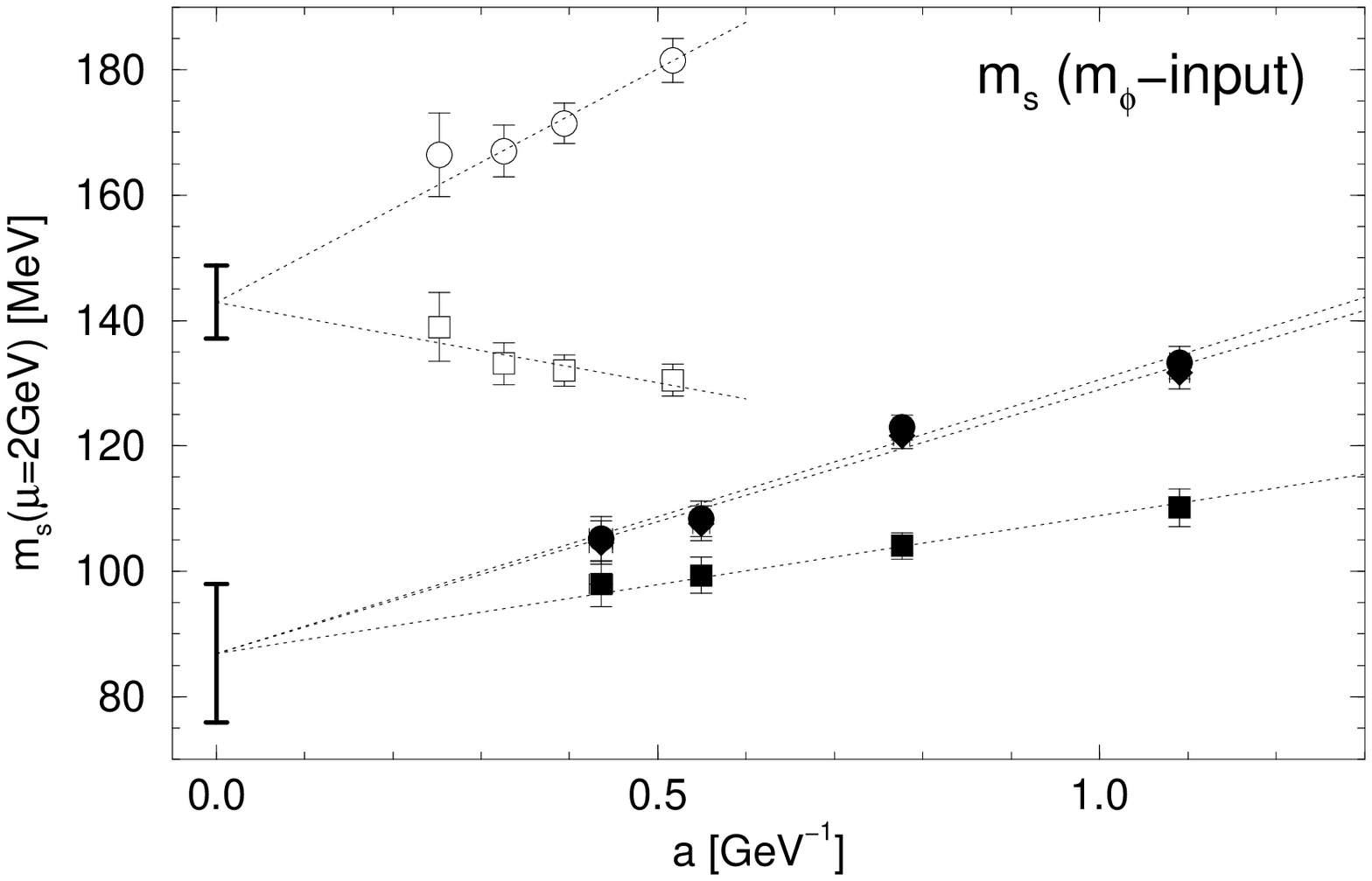}
\end{center}
\vspace{-15mm}
\caption{
Continuum extrapolations of $m_{ud}$ (top), and $m_s$ with
$m_{K}$-input (middle) and $m_{\phi}$-input (bottom) [5].
Open and filled symbols are results for $N_f=0$ and $2$.
}
\vspace{-3mm}
\end{figure}

Here is a summary of results presented by various groups at LAT99.\\
$\bullet$ {\sl CP-PACS:} RG-improved gauge field action and tadpole-improved
clover fermion action are used. Ward identities are used to extract $m_q$.
$N_f=0$ and $N_f=2$ results are compared, demonstrating that going to $N_f=2$
reduces the systematic error in the lattice scale. The results converted to
the $\overline{MS}$ scheme are \cite{cppacs}:
\begin{equation}
\overline{m}_{ud} (2~{\rm GeV}) = 3.3(4)~{\rm MeV} ~,~
\overline{m}_s (2~{\rm GeV}) = 84(7)~{\rm MeV} ~.
\end{equation}
$\bullet$ {\sl ALPHA/UKQCD:} Schr\"odinger functional method is used with
$O(a)$ improved action and $N_f=0$. PCAC relation is used to extract $m_q$.
With $(M_Kr_0)^2=1.5736$ fixing the reference scale, the results are
\cite{alpha}:
\begin{equation}
2M_{ref} = m_s^{RGI} + m_{ud}^{RGI} = 143(5)~{\rm MeV} ~,~
\overline{m}_s (2~{\rm GeV}) = 94(4)~{\rm MeV} ~.
\end{equation}
$\bullet$ {\sl QCDSF:} Schr\"odinger functional method is used with $O(a)$
improved action and $N_f=0$. PCAC relation is used to extract $m_q$. Using
$r_0=0.5~{\rm fm}$ as the reference scale yields \cite{qcdsf}:
\begin{equation}
\overline{m}_{ud} (2~{\rm GeV}) = 4.4(2)~{\rm MeV} ~,~
\overline{m}_s (2~{\rm GeV}) = 105(4)~{\rm MeV} ~.
\end{equation}
$\bullet$ {\sl Rome group:} $O(a)$ improved action is used with $N_f=0$.
Ward identity for the renormalised quark propagator is used to extract $m_q$.
With pseudoscalar and vector meson masses fixing the lattice scale, the
results are \cite{rome}:
\begin{equation}
\overline{m}_{ud} (2~{\rm GeV}) = 4.8(5)~{\rm MeV} ~,~
\overline{m}_s (2~{\rm GeV}) = 111(9)~{\rm MeV} ~.
\end{equation}

Numerical results show that, for a fixed value of the lattice scale,
increasing $N_f$ increases the coupling $g^2$ through vacuum polarisation.
This translates into $m_q(\mu)$ decreasing with increasing $N_f$.
My educated guess, including the effect of a relatively light $s-$quark, is
\begin{equation}
\overline{m}_s (2~{\rm GeV},N_f=2.5) ~=~ 90 \pm 10~{\rm MeV} ~.
\end{equation}

\begin{figure}[t]
\epsfxsize=7.5cm
\centerline{\epsfbox{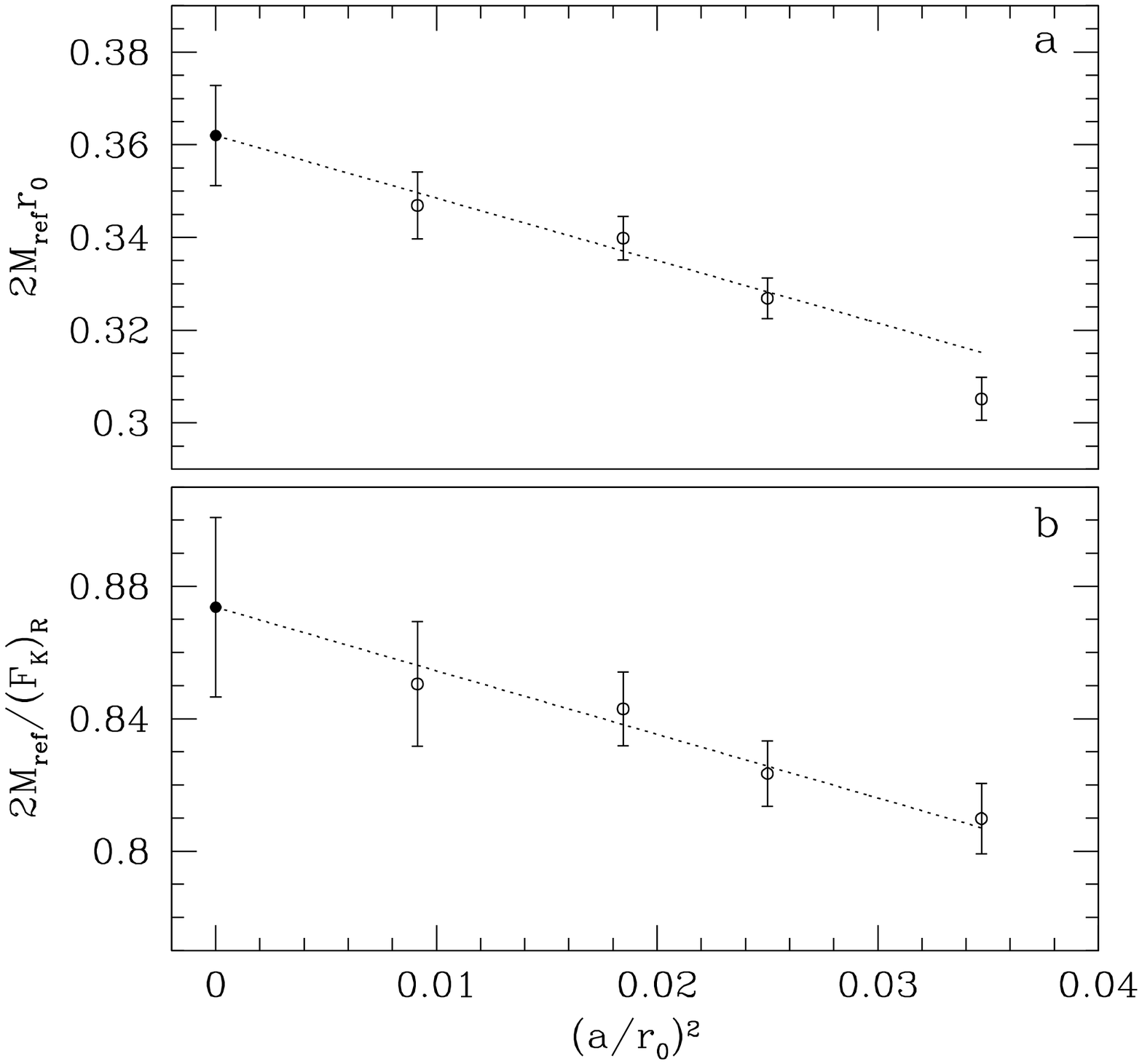}}
\caption{\label{ContPlot}
Continuum extrapolations of $M_{ref}$ in units of (a) $r_0$, and
(b) $(F_K)_{\rm R}$ [6].
The largest value of $a$ is not used in the (dashed) extrapolation to $a=0$.}
\vspace{-3mm}
\end{figure}

\vspace*{-1mm}
\section{Determination of $\epsilon'/\epsilon$}

$\epsilon$ and $\epsilon'$ parametrise indirect and direct $CP-$violation
effects in neutral kaon decays into two pions. $\epsilon$ arises from the
mixing between $CP-$eigenstates $K_L$ and $K_S$, and is experimentally
measured to be
\begin{equation}
\epsilon = 2.280(13) \cdot 10^{-3} ~e^{i\phi_\epsilon} ~,~
\phi_\epsilon \approx \pi/4 ~.
\end{equation}
$\epsilon'$ parametrises $CP-$violation in the decay amplitudes without
$K_L-K_S$ mixing, and in the standard model it arises from the complex
phase in the CKM quark mixing matrix. $\pi\pi-$ scattering data establish
that the phases of $\epsilon$ and $\epsilon'$ are almost identical. Recent
results from NA31, E731, KTeV and NA48 experiments have determined
(see Ref.\cite{buras} for details)
\begin{equation}
Re(\epsilon'/\epsilon) ~=~ (21.2 \pm 4.6) \cdot 10^{-4} ~.
\end{equation}

\vspace*{-1mm}
\subsection{Parametrisation using OPE}

All phenomenological explanations of $\epsilon'/\epsilon$ take the
$CP-$conserving data from experiments and estimate the $CP-$violating
part. In the standard model, $\epsilon$ is found by calculating the box
diagram, while $\epsilon'$ is found by calculating the $g,\gamma,Z^0$
penguin diagrams. Operator mixing and RG-evolution are crucial in the
analysis, and electroweak contributions have become important due to
large $m_t$. The dominant components are the $K\rightarrow\pi\pi$ matrix
elements of the $4-$fermion operators $Q_6$ and $Q_8$, in the $\Delta I=1/2$
and $\Delta I=3/2$ channels. Let $B_{i,\Delta I}$ denote the ratios
of the actual matrix elements to their values in the vacuum saturation
approximation (VSA). Using commonly accepted scale parameters, and including
isospin breaking effects with $\Omega_{\eta+\eta'}=0.25(8)$, an approximate
formula is \cite{buras}
\begin{eqnarray}
{\epsilon' \over \epsilon} ~\approx~ & 13 \cdot Im(V_{td} V_{ts}^*)
      \cdot \left( {110~{\rm MeV} \over \overline{m}_s(2~{\rm GeV})} \right)^2
      \cdot \left( {\Lambda_{\overline{MS}}^{(4)} \over 340~{\rm MeV}} \right)
      \cdot \nonumber \\
& \cdot \left[ B_{6,(1/2)}(1-\Omega_{\eta+\eta'}) - 0.4B_{8,(3/2)} \right] ~.
\end{eqnarray}
I want to emphasise that the appearance of $m_s$ in the above formula is
spurious; it arises because the VSA matrix element values are accompanied by
quark mass factors in the chiral limit. It is silly to express one unknown
matrix element in terms of two other unknowns, $m_s$ and $B_{i,\Delta I}$,
but that has become commonplace. To avoid unwanted systematic errors, one
should therefore either (a) calculate the matrix elements directly, or
(b) use the same calculational framework to evaluate both $m_s$ and the
$B_{i,\Delta I}$. Much of the confusion in the literature has arisen from
not following this simple guideline consistently.

$\epsilon'/\epsilon$ has been evaluated in three different frameworks:
lattice QCD, large$-N_c$ approximation, and chiral quark model. Lattice QCD
simulations show that $B_{8,(3/2)}$ is suppressed below $1$, but provide no
clean result for $B_{6,(1/2)}$. With $B_{6,(1/2)}=1.0(3)$ and
$B_{8,(3/2)}=0.8(2)$, the estimates for $\epsilon'/\epsilon$ are about a
factor $2$ below its experimental value.

\vspace*{-1mm}
\subsection{Final state interactions}

Final state interactions (FSI) strongly influence $K\rightarrow\pi\pi$ decays,
as shown by the experimental phase shifts. They are absent in VSA, lattice
QCD, large$-N_c$ approximation, and appear at subleading order in chiral
perturbation theory. Their contribution to the decay amplitudes can be
estimated using dispersion relation analysis of the experimentally measured
phase shifts in the elastic $\pi\pi$ channel. There is no doubt that
incorporating the FSI boosts the $\Delta I=1/2$ amplitude and suppresses the
$\Delta I=3/2$ one, and hence increases $\epsilon'/\epsilon$. The exact value
of the enhancement is debatable, since it depends on the boundary conditions
used to analyse the experimental data. In a specific analysis \cite{pallante},
the enhancement is the required factor of $2$. Higher order chiral quark
model calculations, which automatically include the FSI, also enhance 
$\epsilon'/\epsilon$ close to its experimental value \cite{trieste}.

In conclusion, the standard model, with a proper framework to handle
non-perturbative QCD effects, is fully capable of explaining the observed
value of $\epsilon'/\epsilon$; contributions to $\epsilon'/\epsilon$ from
new effects beyond the standard model must be kept $\le 5 \cdot 10^{-4}$.

\end{document}